\documentclass[prd,twocolumn,preprint numbers, aps,nofootinbib,showpacs]{revtex4-1}
\usepackage{graphicx}
\usepackage{epsfig}
\usepackage{amsmath}
\usepackage{amsfonts}
\usepackage{amssymb}
\usepackage{url}
\usepackage{hyperref}
\usepackage{subfigure}
\usepackage{hhline}
\usepackage{color}
\usepackage{bm}
\usepackage{cancel}
\usepackage{times}
\newcommand{\beqa}{\begin{eqnarray}}
\newcommand{\eeqa}{\end{eqnarray}}
\newcommand{\beq}{\begin{equation}}
\newcommand{\eeq}{\end{equation}}

\newcommand{\bmt}{\begin{pmatrix}}
\newcommand{\emt}{\end{pmatrix}}
\usepackage[toc,page]{appendix}
\usepackage{comment}
\newcommand{\be}{\begin{equation}}
\newcommand{\ee}{\end{equation}}
\newcommand{\bea}{\begin{eqnarray}}
\newcommand{\eea}{\end{eqnarray}}

\begin{document}

\title{Distinguishing Non-Standard Interaction and Lorentz Invariance Violation at Protvino to Super-ORCA experiment} 

\author{Rudra Majhi$^{1}$}
\email{rudra.majhi95@gmail.com}

\author{Dinesh Kumar Singha$^{1}$}
\email{dinesh.sin.187@gmail.com}

\author{Monojit Ghosh$^{2}$}
\email{mghosh@irb.hr}

\author{Rukmani Mohanta$^{1}$}
\email{rmsp@uohyd.ac.in}

\affiliation{$^1$\,School of Physics, University of Hyderabad, Hyderabad - 500046, India}  
\affiliation{$^2$\,Center of Excellence for Advanced Materials and Sensing Devices, Ruder Bo\v{s}kovi\'c Institute, 10000 Zagreb, Croatia}

\begin{abstract}
As the two phenomena, non-standard interaction (NSI) in neutrino propagation and Lorentz invariance violation (LIV) modify the Hamiltonian of neutrino oscillation in a similar fashion, it is very difficult to distinguish these two effects. The only difference between them lies in the fact that NSI depends on the matter density, whereas LIV is independent of the earth matter effect.
Therefore for a fixed baseline experiment, where matter density is constant, the theories describing NSI and LIV are exactly equivalent. However, as the present and future bounds of the NSI and LIV parameters are not equivalent, one can  distinguish these two scenarios in the long-baseline neutrino experiments depending on their statistics with respect to the present and future bounds of these parameters. In this paper, we attempt to differentiate between LIV and NSI in the context of DUNE and P2SO, as these two future experiments are believed to be sensitive to the strongest matter effect and will have very large statistics. Taking LIV in the data and NSI in theory, our results show that, indeed it is possible to have good discrimination between LIV and NSI. The best separation between LIV and NSI at $3 \sigma$ C.L. is achieved for the parameter $a_{\mu\mu}$ with P2SO. In this case, the value of LIV parameter for which separation is possible, lies within its future bound, if one considers the value of NSI parameter to be constrained by the present experiments. Between DUNE and P2SO, the latter has better sensitivity for such discrimination.

\end{abstract}

\maketitle

\section{Introduction}

The Standard Model (SM) of particle physics is quite successful in explaining the interaction of the elementary particles. However, it is well known that SM can not  unravel some of the  experimental observations, e.g., tiny neutrino masses and neutrino oscillation. Hence, neutrino oscillation opens up the window for  Beyond the SM (BSM) physics. Various neutrino oscillation experiments like solar, atmospheric, reactor and accelerator based have measured the oscillation parameters with great precision within the standard three flavour framework~\cite{Esteban:2020cvm}. Thus, the phenomenon of neutrino oscillation also provides a great opportunity to explore several new physics scenarios beyond the standard three flavour framework.

Two of these new physics scenarios are non-standard interactions (NSI) in neutrino propagation~\cite{Ohlsson:2012kf, Miranda:2015dra, Farzan:2017xzy} and Lorentz invariance violation (LIV)~\cite{Mewes:2004wp} which are studied extensively in neutrino oscillation experiments.  In case of  non-standard interactions, the initial and final flavour of the neutrinos can be different due to the interactions of  the propagating neutrinos  with the earth matter~\cite{Wolfenstein:1977ue}. On the other hand,  it has been shown that Lorentz symmetry which is one of the fundamental symmetries of quantum field theory related to space and time, can be violated in the low energy theories and hence, can be detected through neutrino oscillation~\cite{Colladay:1998fq}. The effect of both NSI and LIV are well explored in the context of the neutrino oscillation. For studies on NSI, we refer to Refs.~\cite{Masud:2015xva,deGouvea:2015ndi,Coloma:2015kiu,Liao:2016hsa,Masud:2016bvp,C:2016nrg,Coloma:2016gei,Masud:2016nuj,Blennow:2016etl,Agarwalla:2016fkh,Blennow:2016jkn,Fukasawa:2016lew,Deepthi:2016erc,Liao:2016orc,Ghosh:2017ged,Masud:2017bcf,Ghosh:2017lim,Deepthi:2017gxg,Meloni:2018xnk,Flores:2018kwk,Verma:2018gwi,Masud:2018pig,Liao:2019qbb,DUNE:2020fgq,Bakhti:2020fde,Chatterjee:2021wac,Feng:2019mno,Singha:2021jkn,Agarwalla:2021zfr,Khatun:2019tad,Kumar:2021lrn,Barranco:2011wx,Barranco:2007ej,Biggio:2009nt,Gonzalez-Garcia:2013usa,Gonzalez-Garcia:2001snt} and for studies in LIV, we refer to Refs.~\cite{Barenboim:2018ctx,KumarAgarwalla:2019gdj,Fiza:2022xfw,Sahoo:2021dit,Super-Kamiokande:2014exs,IceCube:2017qyp,Majhi:2019tfi,Lin:2021cst,Rahaman:2021leu,Crivellin:2020oov,MINOS:2010kat,Coleman:1998ti,Borges:2022kle,Moura:2022dev,PierreAuger:2021tog}. Although the physics behind NSI and LIV are quite different, the modification in the Hamiltonian of neutrino propagation due to NSI and LIV are very similar. The only difference lies in the fact that the effect of NSI is more prominent in the presence of matter whereas LIV is not affected by the matter density. Therefore, in neutrino oscillation experiments, it is arduous to distinguish the effects of NSI and LIV. Recently, an effort was made to distinguish these two phenomena in the context of atmospheric neutrino experiment ICAL~\cite{Sahoo:2022rns} where one can have different earth matter densities over various baseline lengths. For long-baseline experiments, where the matter density is constant, the theories of NSI and LIV are exactly equivalent. However, as the present and future bounds of the NSI and LIV parameters are not in the equal footing, it is possible to discriminate these two scenarios in the long-baseline experiments. Depending on the value of matter density and statistics, one can have a significant difference between the two theories. In this paper, for the first time we study the possibility of distinguishing NSI and LIV at the following long-baseline neutrino experiments: Deep Underground Neutrino Experiment (DUNE)~\cite{DUNE:2020ypp} in Fermilab with a baseline of 1300 km and Protvino to Super-ORCA (P2SO)~\cite{Akindinov:2019flp} at KM3NeT~\cite{KM3Net:2016zxf} facility with a baseline of 2595 km. These two future experiments will have the most notable matter effect and higher statistics.
In our work, our strategy will be to take LIV in the simulated data and NSI in theory and study the capability of these two future neutrino oscillation experiments to distinguish LIV from NSI, using current and future bounds of the NSI parameters. As future experiments will be able to put stronger bound on the NSI parameters, we expect a higher distinguishability of LIV from NSI when we use the future bounds of the NSI parameters in the analysis.

The paper is organized as follows. In the next section, we present the theoretical background of NSI and LIV and illustrate how the Hamiltonian of the neutrino oscillation gets modified by the presence of NSI and LIV. Then we discuss the specification of the experiments and the details of numerical analysis which has been used in our calculation. After that, we will briefly outline the present and future possible bounds on the NSI and LIV parameters. Then we will estimate the sensitivity of DUNE and P2SO to discriminate between the two phenomena with respect to their present and future possible bounds. Finally,  we  summarize our results and conclude. 

\section{Theoretical Background}

The neutral current non-standard interaction between the propagating neutrinos and fermions in the earth matter  can be written in terms of the interaction Lagrangian \cite{Farzan:2017xzy}
\begin{eqnarray}
\mathcal{L}_{NSI} = -2\sqrt{2} G_{F} \epsilon^{f C}_{\alpha \beta} (\overline{\nu}_{\alpha} \gamma^{\mu} P_{L} \nu_{\beta}) (\overline{f} \gamma_{\mu} P_{C} f),
\end{eqnarray}
where $G_F$ represents the Fermi constant, $\epsilon^{f C}_{\alpha \beta}$ are the NSI parameters characterizing the strength of non-standard interactions with $\alpha$, $\beta$ = $e, \mu, \tau$, the fermion fields are represented by $f$ = $e, u, d$, and $C$ = $L, R$ stands for the left  and right chiral projection operators. For the neutrino propagation in the earth, the relevant combinations of the NSI parameters are expressed as
\be
\epsilon_{\alpha \beta}=\sum_{f=e,u,d}\epsilon_{\alpha\beta} \frac{N_f}{N_e}=\sum_{f=e,u,d}\left (\epsilon_{\alpha\beta}^{fL}+\epsilon_{\alpha\beta}^{fR}\right )  \frac{N_f}{N_e}\;,
\ee
where $N_f$ represents the number density of $f$ fermion. For the Earth matter, which is considered as neutral and isoscalar, one can have 
$N_n \simeq N_p=N_e$,  which essentially implies
$N_u\simeq N_d \simeq 3N_e$. Hence, one can write the NSI parameters as
\bea
\epsilon_{\alpha \beta}\simeq \epsilon_{\alpha \beta}^e+3\epsilon_{\alpha \beta}^u+3\epsilon_{\alpha \beta}^d\;.
\eea
The effective Lagrangian density describing Lorentz invariance violation (LIV) in neutrino interactions can be written as \cite{Majhi:2019tfi, Kostelecky:2003cr, Kostelecky:2011gq}
\begin{eqnarray}
\mathcal{L}_{LIV} &=& -\frac{1}{2} [ p^{\mu}_{\alpha \beta} \bar{\nu}_{\alpha} \gamma_{\mu} \nu_{\beta} + q^{\mu}_{\alpha \beta} \overline{\nu}_{\alpha} \gamma_{5} \gamma_{\mu} \nu_{\beta} - \\ \nonumber
&& i r^{\mu \nu}_{\alpha \beta} \bar{\nu}_{\alpha} \gamma_{\mu} \partial_{\nu} \nu_{\beta} - i s^{\mu \nu}_{\alpha \beta} \bar{\nu}_{\alpha} \gamma_{5} \gamma_{\mu} \partial_{\nu} \nu_{\beta} ],
\end{eqnarray}
where $p^{\mu}_{\alpha \beta}$, $q^{\mu}_{\alpha \beta}$, $r^{\mu \nu}_{\alpha \beta}$ and $s^{\mu \nu}_{\alpha \beta}$ are the Lorentz violating parameters in the flavor basis. As only the left handed neutrinos exist in nature,  the LIV parameters can be parameterized by the following observable quantities, which are combinations of  $p^{\mu}_{\alpha \beta}$, $q^{\mu}_{\alpha \beta}$, $r^{\mu \nu}_{\alpha \beta}$ and $s^{\mu \nu}_{\alpha \beta}$:
\begin{eqnarray}
(a_{L})^{\mu}_{\alpha \beta} = (p + q)^{\mu}_{\alpha \beta},~~ {\rm and} ~~ (c_{L})^{\mu \nu}_{\alpha \beta} = (r + s)^{\mu \nu}_{\alpha \beta}\;,
\end{eqnarray}
where $(a_{L})^{\mu}_{\alpha \beta}$ are related to CPT violating neutrino interactions and $(c_{L})^{\mu \nu}_{\alpha \beta}$ are associated with CPT even, Lorentz violating interactions.
For long baseline experiments the effective Hamiltonian in the presence of both NSI and LIV can be written as
\begin{eqnarray}
H = H_{vac} + H_{mat} + H_{NSI} + H_{LIV},
\end{eqnarray}
 where $H_{vac}$ and $H_{mat}$ represent the vacuum and standard matter Hamiltonian whereas $H_{NSI}$ and $H_{LIV}$ represent the NSI and LIV Hamiltonian. The different components of the Hamiltonian can be expressed as
 
 \begin{eqnarray}
 H_{vac} &=& \frac{1}{2 E} U \left(\begin{array}{ccc}
	m^{\rm 2}_{1}&0&0\\
	0&m^{\rm 2}_{2}&0\\
	0&0&m^{\rm 2}_{3}
	\end{array}\right) U^\dagger,~~ \\
H_{mat} &=& \sqrt{2} G_{F} N_{e} \left(\begin{array}{ccc}
	1&0&0\\
	0&0&0\\
	0&0&0
	\end{array}\right),	\\
 H_{NSI} &=& \sqrt{2} G_{F} N_{e} \left(\begin{array}{ccc}
	\epsilon_{e e}&\epsilon_{e \mu}&\epsilon_{e \tau}\\
	\epsilon^{\rm *}_{e \mu}&\epsilon_{\mu \mu}&\epsilon_{\mu \tau}\\
	\epsilon^{\rm *}_{e \tau}&\epsilon^{\rm *}_{\mu \tau}&\epsilon_{\tau \tau}
	\end{array}\right)~~~ {\rm and} ~~ \\
H_{LIV} &=& \left(\begin{array}{ccc}
	a_{e e}&a_{e \mu}&a_{e \tau}\\
	a^{\rm *}_{e \mu}&a_{\mu \mu}&a_{\mu \tau}\\
	a^{\rm *}_{e \tau}&a^{\rm *}_{\mu \tau}&a_{\tau \tau} 
	\end{array}\right) - \frac{4}{3}E \left(\begin{array}{ccc}
	c_{e e}&c_{e \mu}&c_{e \tau}\\
	c^{\rm *}_{e \mu}&c_{\mu \mu}&c_{\mu \tau}\\
	c^{\rm *}_{e \tau}&c^{\rm *}_{\mu \tau}&c_{\tau \tau} 
	\end{array}\right),\hspace*{0.6 true cm}
\label{nsi-liv-eqn}
 \end{eqnarray}
  where  the parameters $m_1$, $m_2$ and $m_3$ represent the masses of the neutrinos, which enter into the neutrino oscillation probabilities as $\Delta m^2_{21} = m_2^2 - m_1^2$ and $\Delta m^2_{31} = m_3^2 - m_1^2$. The PMNS matrix $U$ contains three mixing angles $\theta_{12}$, $\theta_{13}$ and $\theta_{23}$ and one Dirac type phase $\delta_{\rm CP}$.
 If we neglect the contribution from the CPT conserving LIV parameters, then the effect of NSI is analogous to LIV as can be seen in Eqn. (\ref{nsi-liv-eqn}). The only difference is that NSI effects explicitly depend on the matter density while  LIV is independent of matter. From the above equations, one can obtain the correlation between the NSI and LIV  parameters as
 \begin{eqnarray}
a_{\alpha \beta} =  \sqrt{2} G_{F} N_{e} \epsilon_{\alpha\beta}\;.
 \end{eqnarray}
The matter potential part can be written as 
\begin{eqnarray}
\sqrt{2} G_{F} N_{e}\simeq 7.5X_e \frac{\rho}{10^{14}{\rm(g/cm^3)}}{\rm eV},
\end{eqnarray}
where $X_e$ is the relative electron number density and for a neutral medium  $X_e=0.5$. The matter density is given by $\rho$ in the units of g/cm$^3$.
From the above two equation, we obtain
\begin{eqnarray}
a_{\alpha \beta} = 3.75 \times \frac{\rho}{\rm(g/cm^3)} \times 10^{-23} ~\epsilon_{\alpha\beta}~{\rm GeV}.
\label{eqv}
 \end{eqnarray}
For long-baseline experiments $\rho$ is approximately constant. Therefore, we understand that, for an upper bound of the NSI parameter with the value $\epsilon_{\alpha\beta}$, it can be distinguished from LIV only if the LIV parameters are larger than $3.75 \times \rho \times 10^{-23} ~\epsilon_{\alpha\beta}$ {\rm GeV}. For DUNE, the value of $\rho$ is 2.848 g/cm$^3$  and for P2SO the value is 2.95 g/cm$^3$. In the next section, we will see that the present and future bounds of the NSI and LIV parameters are not connected by the above relation and therefore, it is possible to separate these two scenarios in DUNE and P2SO with respect to their current/future bounds. In this case, the significance of the separation will depend on the statistics of both the experiments.

\section{Experimental Setup and Simulation details}

We have used GLoBES~\cite{Huber:2004ka,Huber:2007ji} software package to simulate DUNE and P2SO. We have further used additional plugins of GLoBES to implement NSI~\cite{Kopp:2006wp} and modified the NSI probability engine accordingly to incorporate LIV.

For simulating the long-baseline experiment P2SO, we use the same configuration as used in Ref.~\cite{Singha:2022btw}.  The Protvino accelerator houses a 1.5 km-diameter U-70 synchrotron that generates 450 KW beam corresponding to 4 $\times 10^{20}$ POT annually for P2SO configuration. The Super-ORCA detector located at a distance of 2595 km from Protvino which is ten times more dense than the ORCA detector, will be used in the P2SO experiment. Three years in neutrino mode and three years in antineutrino mode made up to a total run-time of six years for P2SO. In P2SO, the energy window for event calculation ranges from 0.2 GeV to 10 GeV.

For DUNE, we have used the official GLoBES files corresponding to the technical design report~\cite{DUNE:2021cuw}. At a distance of 1300 km from the neutrino source at Fermilab, a 40 kt liquid argon time-projection chamber detector with a power of 1.2 MW is placed. The total run-time for DUNE is 7 years comprising of 3.5 years in neutrino mode and 3.5 years in antineutrino mode, corresponding to $1.1 \times 10^{21}$ POT annually. 

For the estimation of the sensitivity, we use the Poisson log-likelihood and assume that it is $\chi^2$-distributed:
\begin{equation}
 \chi^2_{{\rm stat}} = 2 \sum_{i=1}^n \bigg[ N^{{\rm test}}_i - N^{{\rm true}}_i - N^{{\rm true}}_i \log\bigg(\frac{N^{{\rm test}}_i}{N^{{\rm true}}_i}\bigg) \bigg]\,,
\end{equation}
where $N^{{\rm test}}$ and $N^{{\rm true}}$ are the number of events in the test and true spectra respectively, and $i$ is the number of energy bins. The systematic is incorporated by the method of pull \cite{Fogli:2002pt,Huber:2002mx}. The best-fit values of the standard oscillation parameters and their $1 \sigma$ ranges are adopted from NuFIT \cite{Esteban:2020cvm}, which are listed in Table \ref{table_sparam}. The NSI and LIV parameters and their current and future bounds are presented in Table \ref{tab:nsi-bound} and Table \ref{tab:liv-bound}, and will be  discussed in detail in the next section. Utilizing the built-in minimizer in GLoBES and incorporating the priors corresponding to their $1 \sigma$ uncertainties, we marginalized over all the relevant parameters in our analysis.
The NSI parameters are varied and marginalized according to Table \ref{tab:nsi-bound}. For $\delta_{C P}$, we have not considered any prior initially and allowed it to vary freely. For systematic uncertainties, we have considered the overall normalization and shape errors corresponding to signal and background. We list the values of systematic errors for P2SO and DUNE in Table \ref{table_sys}. It should be noted that the DUNE GLoBES file contains no shape error.  We show all our results for the normal hierarchy of the neutrino masses i.e., for $\Delta m^2_{31} > 0$.

\begin{table} 
\centering
\begin{tabular}{|c|c|} \hline
Parameters            & True values $\pm$ $1\sigma$       \\ \hline
$\sin^2 \theta_{12}$  & $0.304^{+ 0.013}_{- 0.012}$      \\ 
$\sin^2 \theta_{13}$ & $0.0222^{+ 0.00068}_{- 0.00062}$                 \\ 
$\sin^2 \theta_{23} $ & $0.573^{+ 0.018}_{- 0.023}$                 \\ 
$\delta_{\rm CP}[^\circ] $  & $ 195^{+ 52}_{- 25}$         \\ 
$\Delta m^2_{21}$ [10$^{-5}$ eV$^2$]    & $7.42^{+ 0.21}_{- 0.20}$  \\ 
$\Delta m^2_{31}$ [10$^{-3}$ eV$^2$]   & $2.515^{+ 0.028}_{- 0.028}$    \\ 

 \hline
\end{tabular}
\caption{The values of oscillation parameters that we considered in our analysis. Standard oscillation parameters considered from~\cite{Esteban:2020cvm} with their corresponding 1$\sigma$ errors.}
\label{table_sparam}
\end{table}   

\begin{table} 
\centering
\begin{tabular}{|c|c|c|} \hline
Systematics     & P2SO          & DUNE  \\ \hline
Sg-norm $\nu_{e}$   & 5$\%$   & 2$\%$      \\ 
Sg-norm $\nu_{\mu}$    & 5$\%$            & 5$\%$ \\ 
Bg-norm    & 12$\%$     & 5$\%$ to 20$\%$\\ 
Sg-shape      & 11$\%$     & -\\ 
Bg-shape     & 4\% to 11$\%$       & - \\ 
\hline
\end{tabular}
\caption{The values of systematic errors that we considered in our analysis. ``norm" stands for normalization error, ``Sg" stands for signal and ``Bg" stands for background.}
\label{table_sys}
\end{table}  

\section{Bounds on NSI and LIV parameters}

\begin{table}[htb]
    \centering
    \begin{tabular}{|c|c|c|}
    \hline
    \multicolumn{3}{|c|}{NSI} \\ 
    \hline
    Parameter & Current Bound    & Future Bound\\ 
    \hline
       $\epsilon_{e\mu} $& $< 0.3$  \cite{Biggio:2009nt}& $[-0.036, +0.034]$ \cite{Chatterjee:2021wac}\\
       \hline  
      $\epsilon_{e\tau}$ & $[-0.19, 0.13]$ \cite{Super-Kamiokande:2011dam} & $[-0.031, +0.031]$ \cite{Chatterjee:2021wac} \\
       \hline 
     $  \epsilon_{ee}$ & $< 4$  \cite{Biggio:2009nt} & $[-0.206, +0.278]$ \cite{Chatterjee:2021wac}\\ 
     \hline 
     $\epsilon_{\mu\mu}$ & $ < 0.068$  \cite{Biggio:2009nt} & $<0.0307$  \cite{Davidson:2003ha}  \\
     \hline
       $\epsilon_{\mu\tau}$ & $ < 0.011 $ \cite{Super-Kamiokande:2011dam}& $[-0.0043, 0.0047]$  \cite{Choubey:2014iia} \\
       \hline
  $\epsilon_{\tau\tau}$ & $[-4.4,3.9]$\cite{Davidson:2003ha}&$[-0.03,0.017]$ \cite{Choubey:2014iia} 
      \\ \hline
    \end{tabular}
    \caption{Current and future bounds on NSI Parameters at 90\% C.L.}
    \label{tab:nsi-bound}
\end{table}

\begin{figure*}[htb]
    \centering
    \includegraphics[scale=0.95]{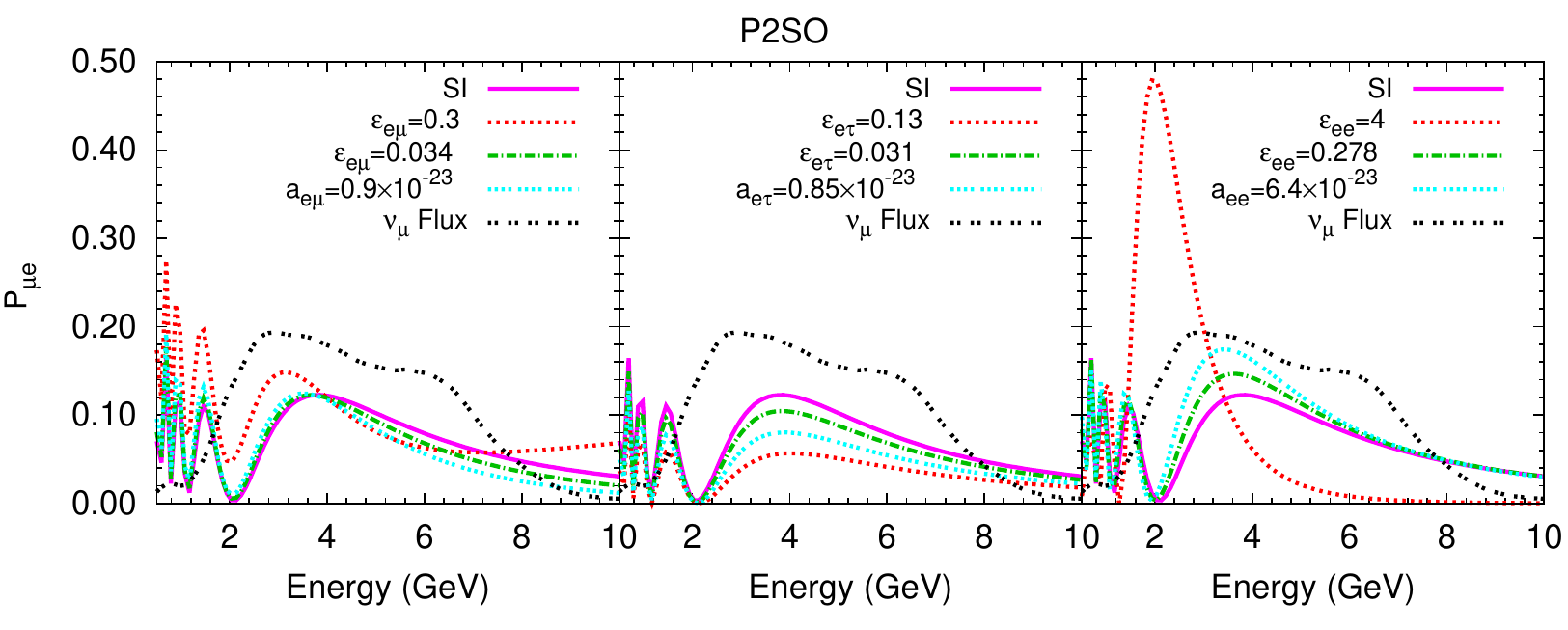}
       \caption{Appearance probability as a function of Neutrino Energy for P2SO experiment in presence of ($\epsilon_{e\mu}$, $a_{e\mu}$), ($\epsilon_{e\tau}$, $a_{e\tau}$) and ($\epsilon_{ee}$, $a_{ee}$) .}
    \label{fig:App-prob}
 \end{figure*}

In Tab.~\ref{tab:nsi-bound}, we have listed the current and future bounds on the NSI parameters at 90\% C.L.. The upper bounds on the NSI parameters are calculated in many studies~\cite{Biggio:2009nt,Farzan:2017xzy,Ohlsson:2012kf,Davidson:2003ha,Proceedings:2019qno, Miranda:2015dra,Gonzalez-Garcia:2013usa}. The bounds on the NSI parameters can be either model dependent or model independent. For our study, we consider only the model independent bounds. The strongest current bounds for $\epsilon_{e\mu}$, $\epsilon_{ee}$ and $\epsilon_{\mu\mu}$ are calculated in Ref.~\cite{Biggio:2009nt} by considering the global data. For the parameters $\epsilon_{e\tau}$ and $\epsilon_{\mu\tau}$, the current strongest bounds come from the Super-Kamiokande data~\cite{Super-Kamiokande:2011dam} whereas for $\epsilon_{\mu\mu}$ a stronger bound has been obtained in the KamLAND+SNO/Super-K data~\cite{Davidson:2003ha}. In future, the strongest bounds for the parameters $\epsilon_{e\mu}$, $\epsilon_{e\tau}$ and $\epsilon_{ee}$ are expected to come from DUNE~\cite{Chatterjee:2021wac} and regarding parameters $\epsilon_{\mu \tau}$ and $\epsilon_{\tau \tau}$ the best future bounds will come from PINGU~\cite{Choubey:2014iia}. For the parameter $\epsilon_{\mu\mu}$, the stringent future bound is expected to come from the neutrino factory experiments~\cite{Davidson:2003ha}. From Table~\ref{tab:nsi-bound}, we understand that future experiments will be able to put an order of magnitude better constraints on the NSI parameters, compared to current bounds except $\epsilon_{\mu \mu}$.

\begin{table}[htb]
    \centering
    \begin{tabular}{|c|c|c|}
    \hline
    \multicolumn{3}{|c|}{LIV} \\
    \hline
    Parameter & Current Bound   & Future Bound  \\
    \hline
       $a_{e\mu}$ & $< 2.4$  \cite{Super-Kamiokande:2014exs}&  $< 0.39$ \cite{Fiza:2022xfw}\\ 
       \hline 
       $a_{e\tau}$ & $< 4.24$ \cite{Super-Kamiokande:2014exs}& $<$0.55 \cite{Fiza:2022xfw}\\ 
       \hline 
       $a_{ee}$ & $[-49.5, +26.5]$\cite{Majhi:2019tfi} &  $[-1.95 , 2.46]$ \cite{Fiza:2022xfw}\\ 
       \hline
$a_{\mu\mu}$ & $[-10, 11]$ \cite{Majhi:2019tfi}& $[-1.24, 1.39]$ \cite{Fiza:2022xfw}\\
      \hline
      $ a_{\mu\tau}$ & $< 0.79$  \cite{Super-Kamiokande:2014exs}& $[-0.19, 0.18]$ \cite{Sahoo:2021dit} \\ 
    \hline
    $a_{\tau\tau}$ & $[-9.8, 8.2]$\cite{Majhi:2019tfi} & $-$  \\
    \hline 
    \end{tabular}
    \caption{Current and future bounds on LIV Parameters in the units of $10^{-23}$ GeV at 90\% C.L.}
    \label{tab:liv-bound}
\end{table}
In Table \ref{tab:liv-bound}, we have shown the bounds on LIV parameters from different neutrino experiments at 90\% C.L. Currently Super-Kamiokande experiment can give strong bounds on $a_{e\mu}$, $a_{e\tau}$ and $a_{\mu\tau}$~\cite{Super-Kamiokande:2014exs}. The current best limits on LIV parameters $a_{ee}$, $a_{\mu\mu}$ and $a_{\tau\tau}$ are obtained from combination of NO$\nu$A and T2K~\cite{Majhi:2019tfi}. In future, very constraint limits on all  parameters except $a_{\mu\tau}$ and $a_{\tau\tau}$ are expected from the combination of  DUNE and P2O as discussed in Ref.~\cite{Fiza:2022xfw}. ICAL experiment can put the strongest bound on $a_{\mu\tau}$ as discussed in Ref.~\cite{Sahoo:2021dit}. To the best of our knowledge, the future model independent bound on the parameter $a_{\tau\tau}$ is yet to be determined. From Table~\ref{tab:liv-bound}, we understand that future experiments will be able to put an order of magnitude better bounds on the LIV parameters, compared to current bounds except $a_{\mu \tau}$.

Note that the bounds on different NSI and LIV parameters are calculated with different initial conditions. For example, with different values of oscillation parameters, inclusion of different number of parameters at a time etc. Therefore, it would be improper to treat all the bounds at the same footing and it only gives us an order of magnitude estimation. 

From Tabs.~\ref{tab:nsi-bound} and \ref{tab:liv-bound}, we see that the present/future bounds of the NSI and LIV parameters do not obey the equivalence relation as given in Eq.~\ref{eqv}. For example, the bound on $\epsilon_{\mu\mu}$, is one order of magnitude stronger than the bound on $a_{\mu\mu}$.   This provides the opportunity to separate these two phenomena in the fixed density experiments with respect to their bounds, which we will see in the next section.

\section{Discrimination at probability levels}

In this section, we will try to distinguish between NSI and LIV effects at the probability level with the long baseline experiments. The neutrino oscillation probabilities are modified in the presence of NSI and LIV.
The probability of finding a neutrino $\nu_\beta$ from a $\alpha$ type neutrino $\nu_\alpha$, after the propagation of distance  $L$ is 
 \bea
 P_{\alpha \beta}= \left | \langle \nu_\beta| \nu_{\alpha}(L)\rangle \right |^2=\left |  \langle \nu_\beta| e^{-i H L} |\nu_{\alpha}\rangle \right |^2, \label{Prob}
\eea 
where $H$ is the effective Hamiltonian in presence of LIV or NSI effects, which we discussed earlier. Exact expression for oscillation probability can be found in Refs. \cite{Kopp:2007ne, Yasuda:2007jp,Masud:2015xva,Masud:2016gcl,Masud:2016bvp,Deepthi:2016erc,Liao:2016hsa,Deepthi:2017gxg,Dey:2018yht,Chatterjee:2018dyd,Chaves:2018sih}.

    \begin{figure*}[htb]
    \centering
    \includegraphics[scale=0.95]{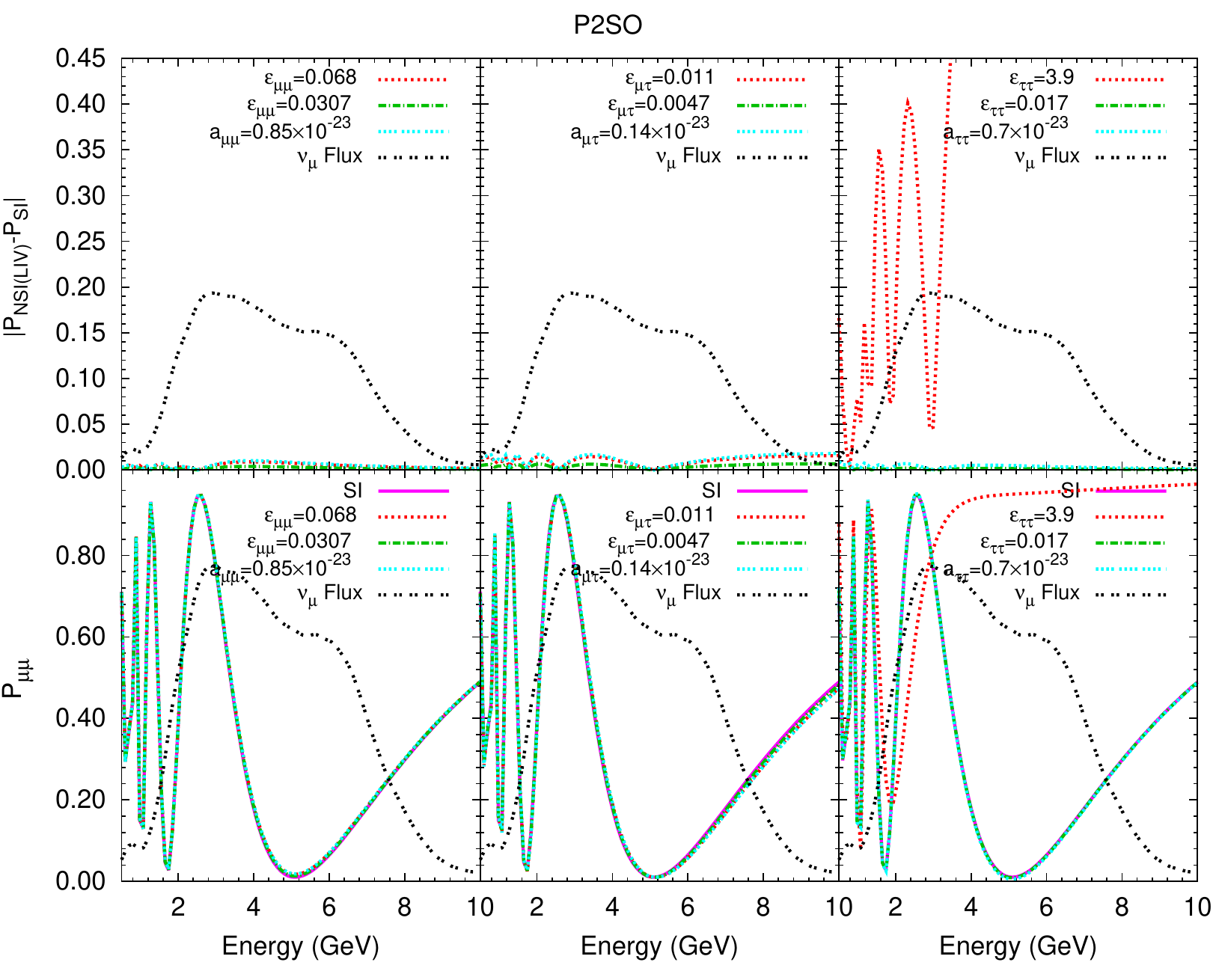}
    \caption{Survival probability  as a function of Neutrino Energy for  P2SO experiment in presence of ($\epsilon_{\mu\mu}$, $a_{\mu\mu}$), ($\epsilon_{\mu\tau}$, $a_{\mu\tau}$) and ($\epsilon_{\tau\tau}$, $a_{\tau\tau}$) .}
    \label{fig:Disapp-prob-p2o}
\end{figure*}
In these calculations, it is shown that the NSI as well as LIV parameters corresponding to $e\mu$, $e\tau$ and $ee$ sectors have leading order contribution to the appearance probability i.e., $\nu_\mu \rightarrow \nu_e$, while the NSI and LIV parameters involving $\mu\mu$, $\mu\tau$ and $\tau\tau$ sectors  affect significantly to  the survival probability $\nu_\mu \rightarrow \nu_\mu$. Therefore, while discussing the separation between NSI and LIV for $e\mu$, $e\tau$ and $ee$ ($\mu\mu$, $\mu\tau$ and $\tau\tau$) sectors, we will use appearance (disappearance) probability only. In our analysis, we have considered one parameter at a time for each NSI and LIV scenarios. Therefore, in our study, we have six independent LIV parameters and six independent NSI parameters. Note that, if we do not consider one parameter at a time and rather consider all the parameters at the same time, then number of independent parameters in each LIV and NSI cases would be five. This is because, when one considers all the parameters at the same time, then one can always subtract a matrix proportional to the identity without changing the oscillation probabilities, which leaves only two of the diagonal parameters to be independent.

First, let us discuss the separation between NSI and LIV effects corresponding to the parameters sensitive to the appearance channel. Fig.~\ref{fig:App-prob} shows the oscillation probability for the $\nu_e$ appearance  channel   as a function of neutrino energy for the P2SO experiment. The left/middle/right panel corresponds to the parameter $e\mu$/$e\tau$/$ee$. In each panel, the standard three flavor case (SI) is represented by the solid magenta curve and black dashed curve represents muon flux of P2SO in arbitrary unit. Therefore, the energy region covered by the black curve shows the region to which the P2SO experiment is sensitive to. In these panels, red (green) dashed curve shows the probability including NSI to SI with the value of the NSI parameters corresponding to their current (future) upper bounds as given Tab.~\ref{tab:nsi-bound}. For the LIV parameter, we choose a value (cyan) for which, a $3\sigma$ separation between LIV and NSI is obtained for P2SO, considering the constraints on the NSI parameters from the future experiments as given in Tab.~\ref{res}.  The first thing we observe from these panels is that,  the first oscillation maxima of P2SO is within the corresponding flux envelope in standard case as well as in NSI and LIV scenarios. We also note that in all the three panels, around the first oscillation maximum, the cyan curve is always sandwiched between the magenta  and red curves but it is mostly outside the region between the magenta and green curves. Or in other words, the difference between SI and NSI is higher (lower) as compared to the difference between SI and LIV for the case of present (future) bound of the NSI parameters. Therefore, it will be very easy to match the probability of LIV with the NSI probability, i.e., the possibility of a degeneracy between NSI and LIV will be more when the NSI parameters are varied within their current upper bounds,  compared to their future bounds. This in turn leads to the fact that, the distinction between NSI and LIV would be higher, when we consider the future bounds of NSI parameters compared to their current bounds,  for a given value of  LIV parameter. Among these three parameters, we expect the separation between NSI and LIV to be very weak for $ee$, if we consider the current bound of the NSI parameter.

Now let us discuss the separation between NSI and LIV corresponding to the parameters, that are sensitive to the disappearance channel, as shown in  Fig.\ref{fig:Disapp-prob-p2o}. The plots in the lower panel are same as Fig.~\ref{fig:App-prob} but for the disappearance channel and for the parameters  $\mu\mu$ (left panel), $\mu\tau$ (middle panel) and $\tau\tau$ (right panel). We have followed the same philosophy as the appearance channel for choosing the values of NSI and LIV parameters i.e., red (green) curve shows the probability including NSI to SI with the values of the NSI parameters corresponding to their current (future) upper bounds. The cyan curve is for the LIV parameter and we choose its value such that, a $3\sigma$ separation between LIV and NSI is obtained for P2SO considering the future bounds of the NSI parameters. As it is difficult to visualize different probabilities at the first minimum, in the upper panel, we have plotted the probability difference between SI to NSI and SI to LIV. From these plots, we also see that the difference between SI and NSI is either higher or equal to SI and LIV, when we consider the values of NSI parameters corresponding to their current bounds and it is lower when we consider the values of NSI corresponding to their future bounds. Therefore,  as in appearance channel, we expect to have better separation between LIV and NSI for future bounds of NSI compared to their current bounds. Among these three parameters, the separation for $\tau\tau$ is found to be very weak, if we consider the current bounds of the NSI parameters.

For DUNE, we have checked that the behaviour of the probability plots are very similar to that of P2SO, and therefore we do not present the figures for DUNE.

\section{Elimination of degeneracy}

\begin{figure*}
    \centering
    \includegraphics[scale=0.95]{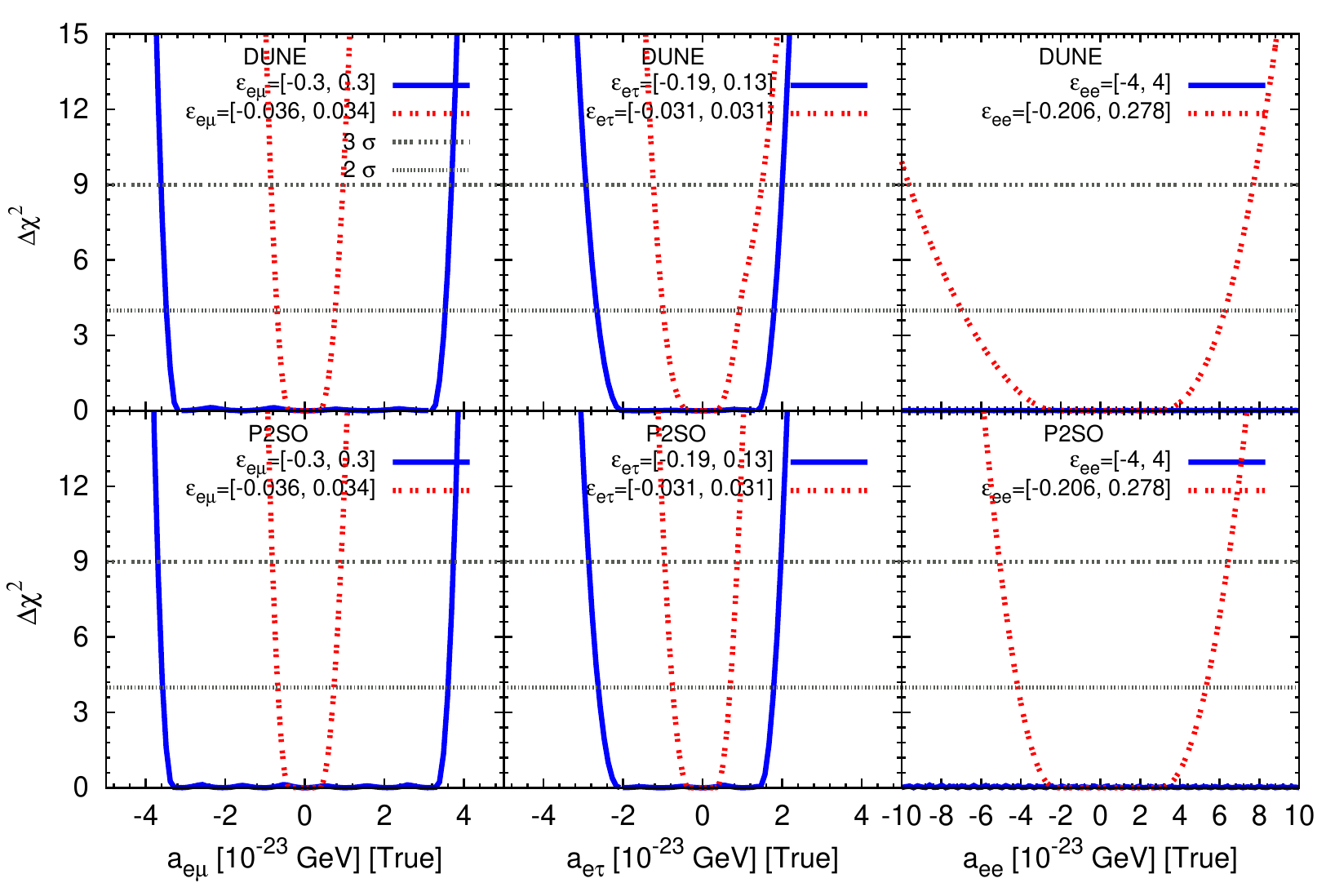}
    \caption{$\Delta{\chi^{2}}$ as a function of true LIV parameters $a_{e\mu}$, $a_{e\tau}$ and $a_{ee}$ for P2SO and DUNE experiments.}
    \label{fig:ee-emu-etau}
\end{figure*}
\begin{figure*}
    \centering
    \includegraphics[scale=0.95]{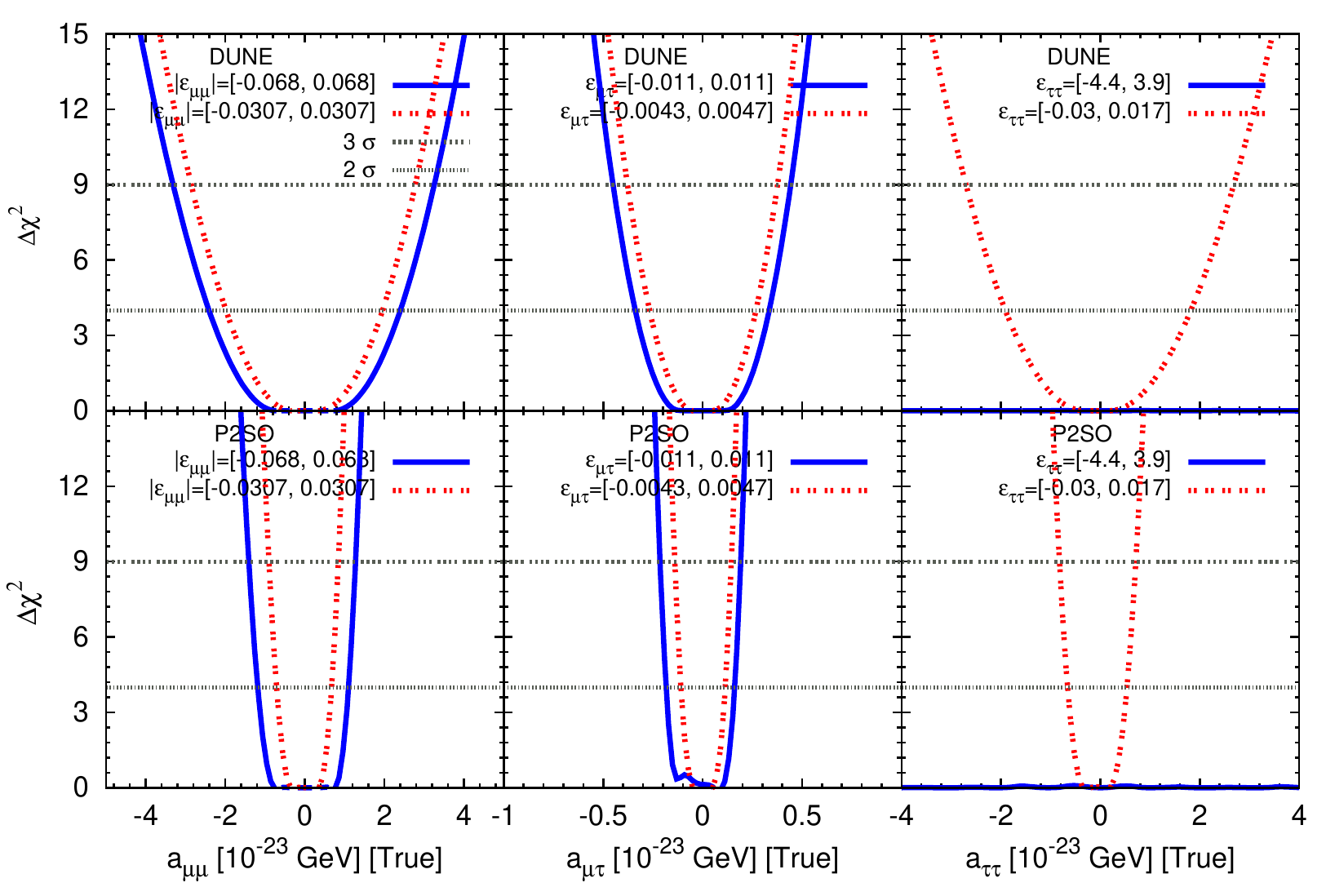}
    \caption{$\Delta{\chi^{2}}$ as a function of true LIV parameters $a_{\mu\mu}$, $a_{\mu\tau}$ and $a_{\tau\tau}$ for P2SO and DUNE experiments.}
    \label{fig:mumu-mutau-tautau}
\end{figure*}

In order to discriminate the degeneracy between NSI and LIV phenomena, we have calculated $\chi^2_{\rm min}$ considering LIV exists in nature and NSI as the test hypothesis, i.e.,
\bea \nonumber
\chi^2 \sim N^{\rm test} (\epsilon^{\rm test}_{\alpha \beta}\neq 0, a^{\rm test}_{\alpha\beta}=0) - N^{\rm true} (\epsilon^{\rm true}_{\alpha \beta}= 0, a^{\rm true}_{\alpha\beta}\neq 0) .
\eea
We have obtained sensitivities for two cases of the NSI parameters i.e., minimizing the NSI parameters either within their current bounds or their future bounds. In Fig.~\ref{fig:ee-emu-etau}, we have shown the sensitivities for the parameters which appear in the appearance channel probabilities i.e., $e\mu$, $e\tau$ and $ee$ and in Fig.~\ref{fig:mumu-mutau-tautau}, we have presented the sensitivities for the parameters  appearing in the disappearance channel probabilities i.e., $\mu\mu$, $\mu\tau$ and $\tau\tau$. In each figure, the plots in the upper panel are for DUNE and in the lower panel are for P2SO. In each panel, the blue curve represents the sensitivity for the current range of the NSI parameters and the red curve represents the sensitivity for the future bounds of the NSI parameters.  The black horizontal lines correspond to $\Delta \chi^2 = 4$ and 9, represent the benchmark sensitivities of $2 \sigma$ and $3 \sigma$, respectively. 

The panels in Figs. \ref{fig:ee-emu-etau} and \ref{fig:mumu-mutau-tautau} can be interpreted in the following way. In these panels, a non-zero value of the $\Delta \chi^2$ signifies the sensitivity to discriminate LIV from NSI. On the other hand, the values of $a_{\alpha \beta}$ for which $\Delta \chi^2 = 0 $, one cannot distinguish LIV from NSI at all. The values of $a_{\alpha \beta}$, for which $\Delta \chi^2$ is grater than 9,  it is possible to separate LIV from NSI at $3 \sigma$ C.L.. From the figures, we see that in general the separation between LIV and NSI is better if we consider future bounds of the NSI parameters, compared to the present limits. This improvement is more in $e\mu$ and $e\tau$ sectors  compared to the $\mu\mu$ and $\mu\tau$ sectors.  For the parameter $ee$ and $\tau\tau$, it is impossible to discriminate LIV from NSI, if we consider the current ranges of the NSI parameters. Between the experiments DUNE and P2SO, better separation between NSI and LIV is obtained for P2SO except the $e\mu$ sector.

In Tab.~\ref{res}, we have listed the values of LIV parameters for which a $3 \sigma$ separation with NSI is possible. 
\begin{table}[htb]
    \centering
    \begin{tabular}{|c|c|c|}
    \hline
    \multicolumn{3}{|c|}{LIV Parameters} \\
    \hline
    Parameter & P2SO   & DUNE  \\
    \hline
       $a_{e\mu}$ &  3.7 (0.9)&  3.7 (0.95)\\ 
       \hline 
       $a_{e\tau}$ & 1.9 (0.85) & 2 (1.5) \\ 
       \hline 
       $a_{ee}$ &  - (6.4) &  - (7.7)\\ 
       \hline
$a_{\mu\mu}$ & 1.3 (0.85)&  3.2 (2.8)\\
      \hline
      $ a_{\mu\tau}$ & 0.2 (0.14) &  0.44 (0.38) \\ 
    \hline 
    $ a_{\tau\tau}$ & - (0.7) &  - (2.65) \\ 
    \hline 
    \end{tabular}
    \caption{Values of LIV parameters in the units of $10^{-23}$ GeV for which a $3 \sigma$ separation with NSI is possible for present (future) bounds on the NSI parameters.}
    \label{res}
\end{table}

 From the table, we understand that the better discrimination between LIV and NSI is possible for the P2SO experiment. Regarding a $3 \sigma$ discrimination between LIV and NSI, from Tables.\ref{res} and \ref{tab:liv-bound}, we can infer the following,  in the context of P2SO:
\begin{itemize}
\item The best discrimination between LIV and NSI is possible for the  $\mu\mu$ sector. For this sector, it is possible to have a separation for the value of LIV parameter {\it within its future bound}, if one considers the value of NSI parameter to be constrained by the {\it present} experiments. 
\item For the sectors $e\tau$ and $\mu\tau$, the discrimination is possible for the values of LIV parameters {\it within their present bounds but outside their future bounds}, if one considers the values of NSI parameters to be constrained by the {\it present} experiments.
\item For $e\mu$ sector, discrimination occurs for the values of LIV parameter {\it within its present bound but outside its future bound}, if one considers the value of the NSI parameter to be constrained by the {\it future} experiments. For values of NSI parameter {\it inside the present} bound, the values of LIV parameter for which discrimination is possible, {\it lies outside the present bound}.
\item Regarding the sectors $ee$ and $\tau\tau$ the sensitivity is worst. For these two cases, the separation is possible for the values of LIV parameters {\it within their present bounds and outside the future bound (for $ee$)}, if one considers the values of NSI parameters to be constrained by the {\it future} experiments. However, no separation is possible if one considers the NSI parameters to be constrained by the present experiments.
\end{itemize}

\begin{figure*}
    \centering
    \includegraphics[scale=0.39]{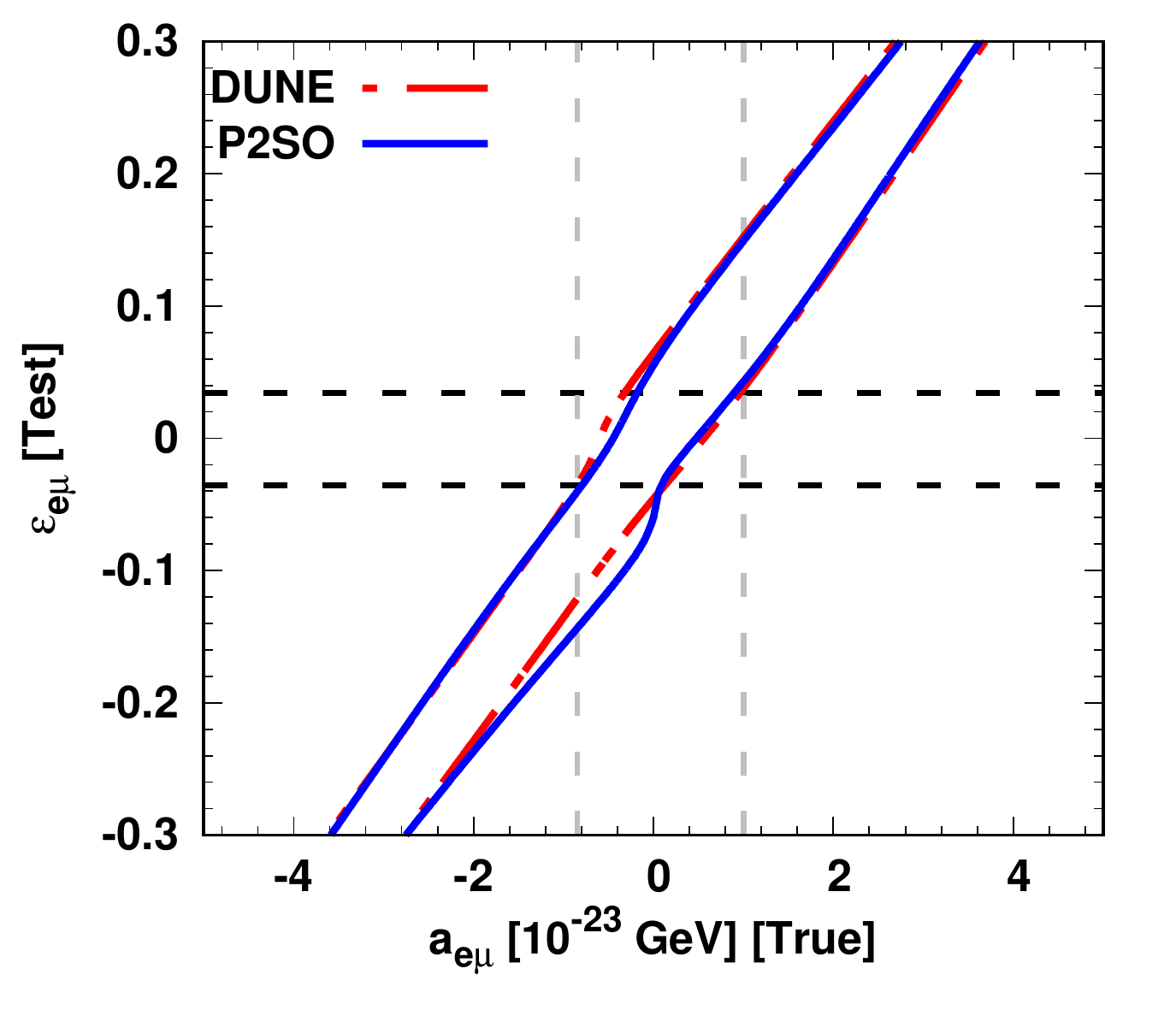}
    \includegraphics[scale=0.39]{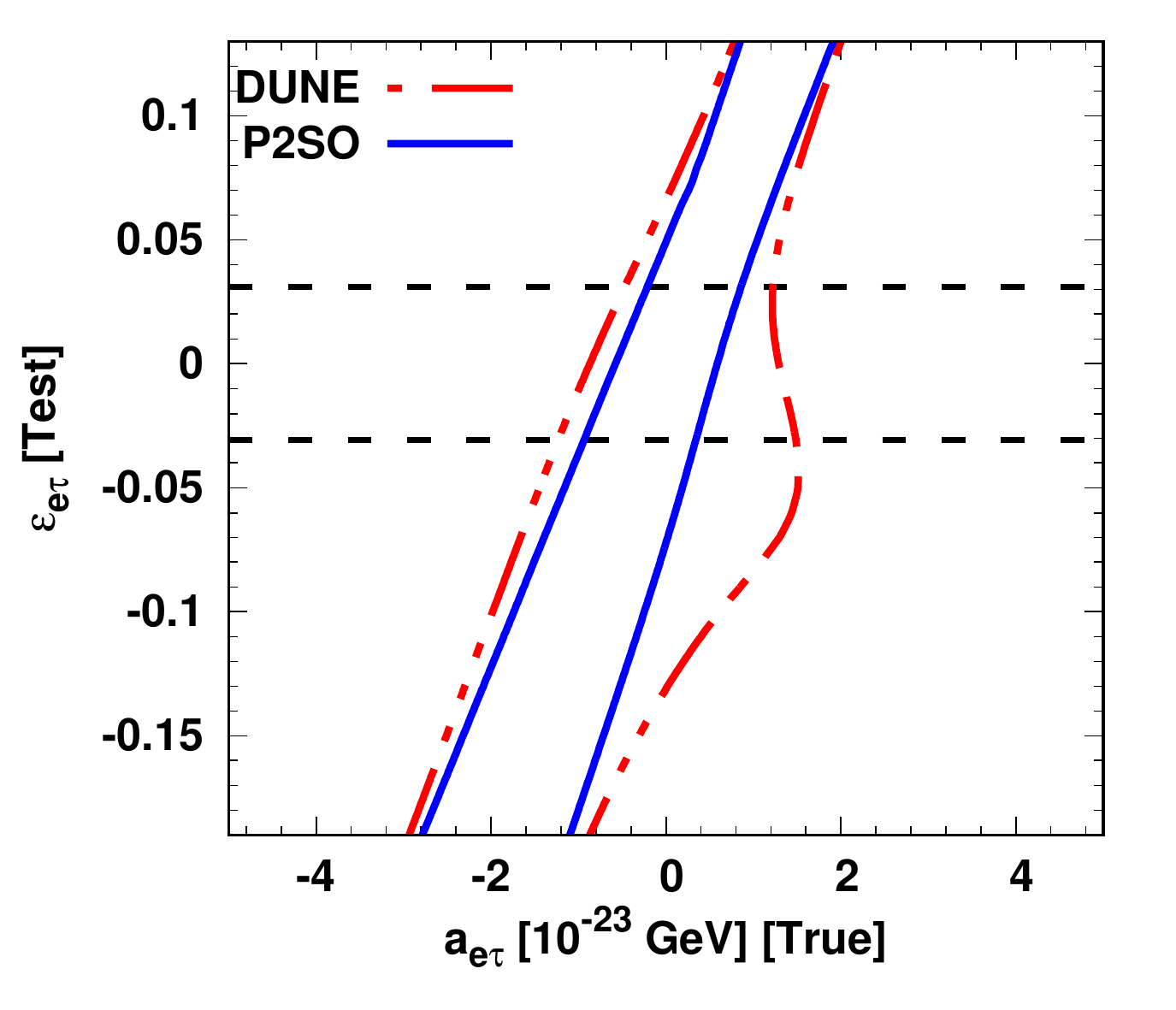}
    \includegraphics[scale=0.39]{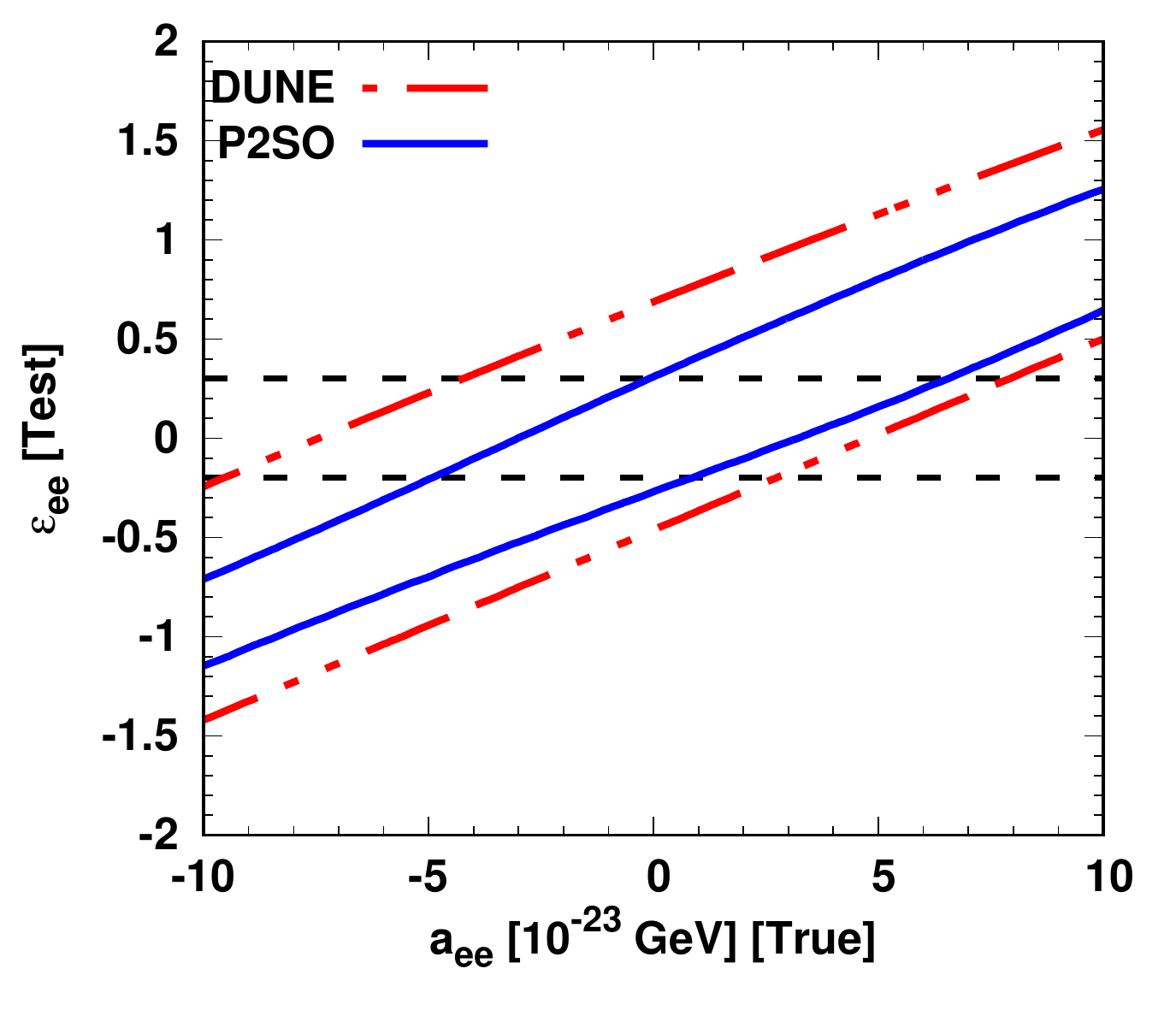}\\
    \includegraphics[scale=0.39]{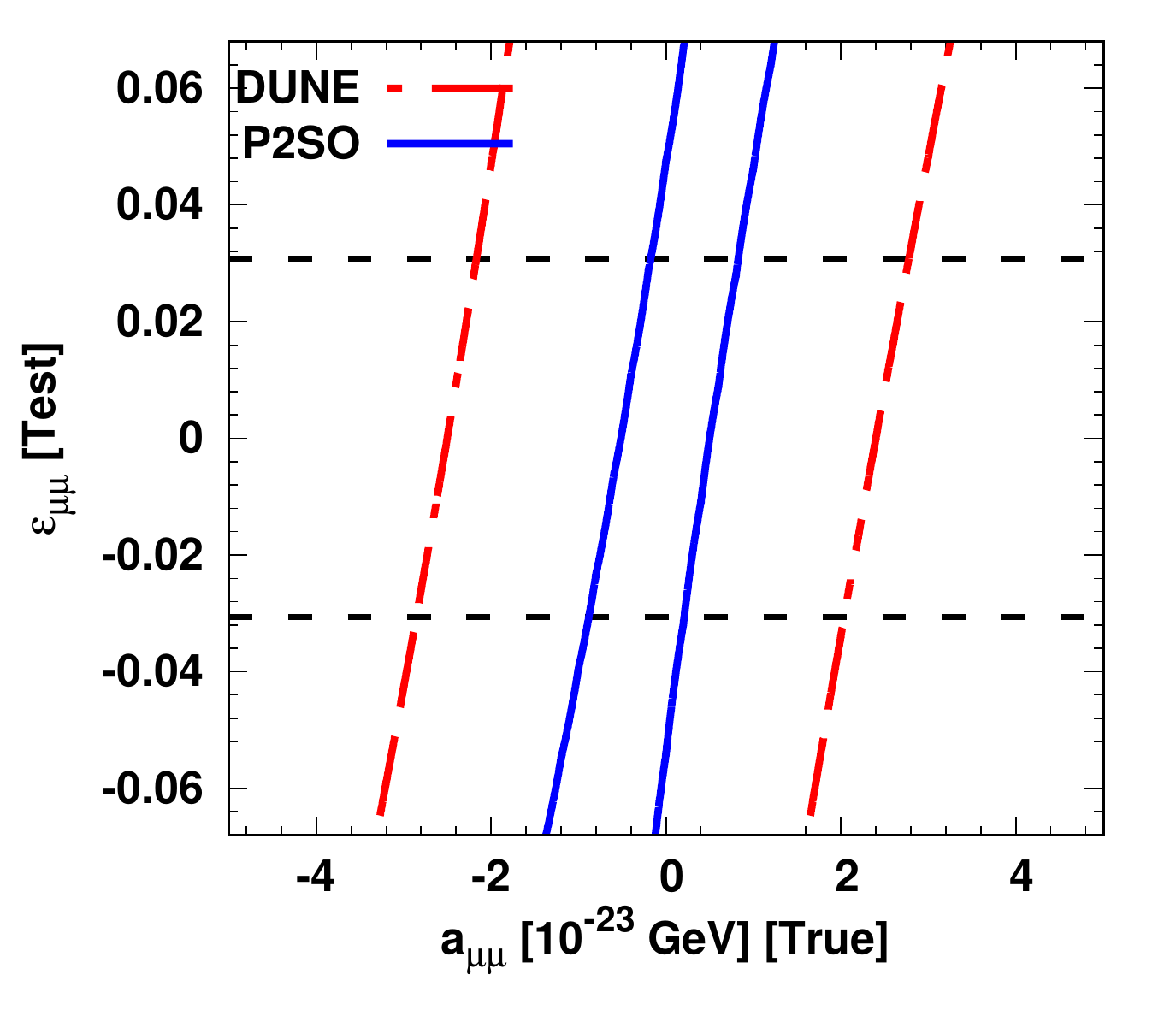}
    \includegraphics[scale=0.39]{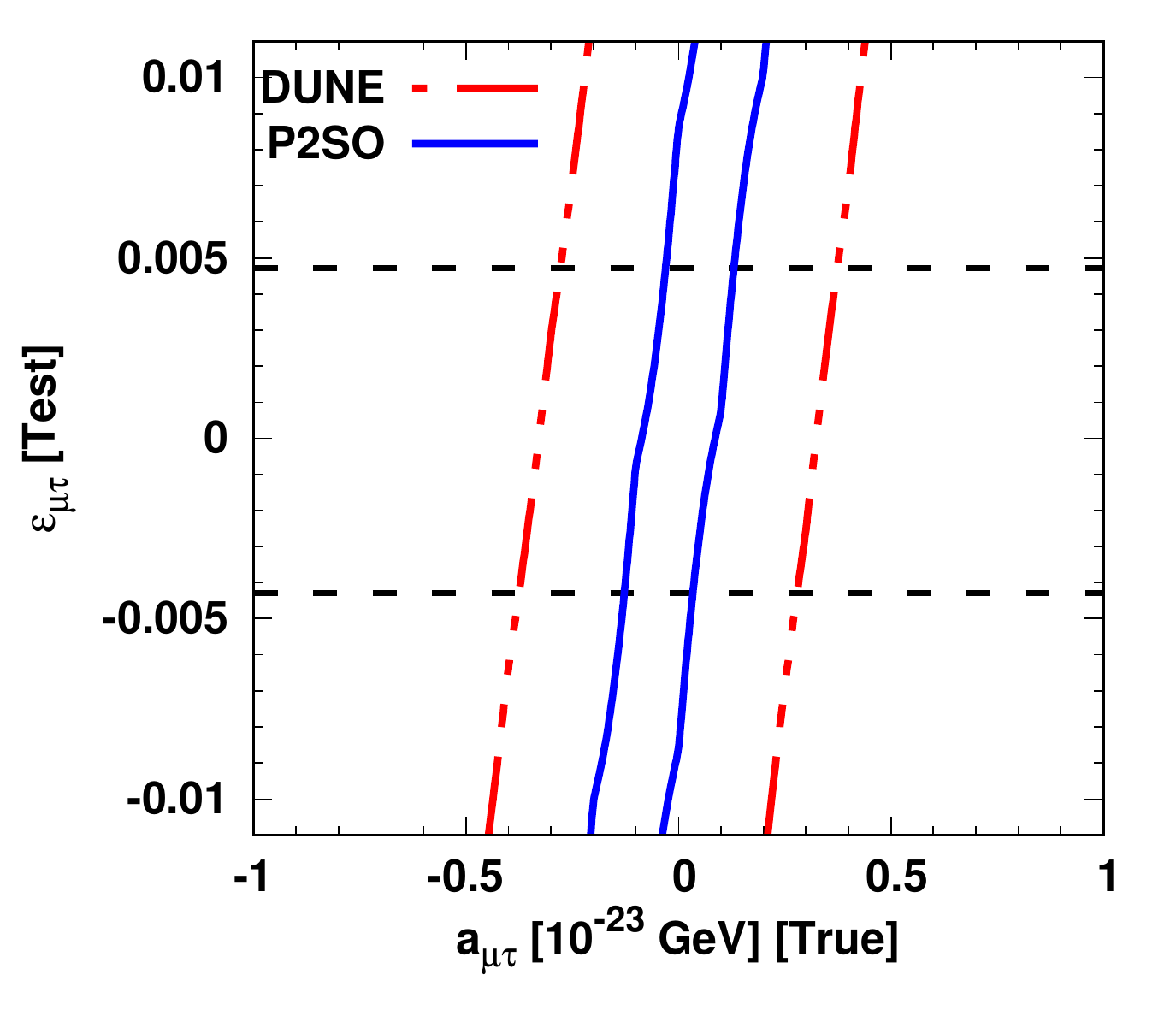}
    \includegraphics[scale=0.39]{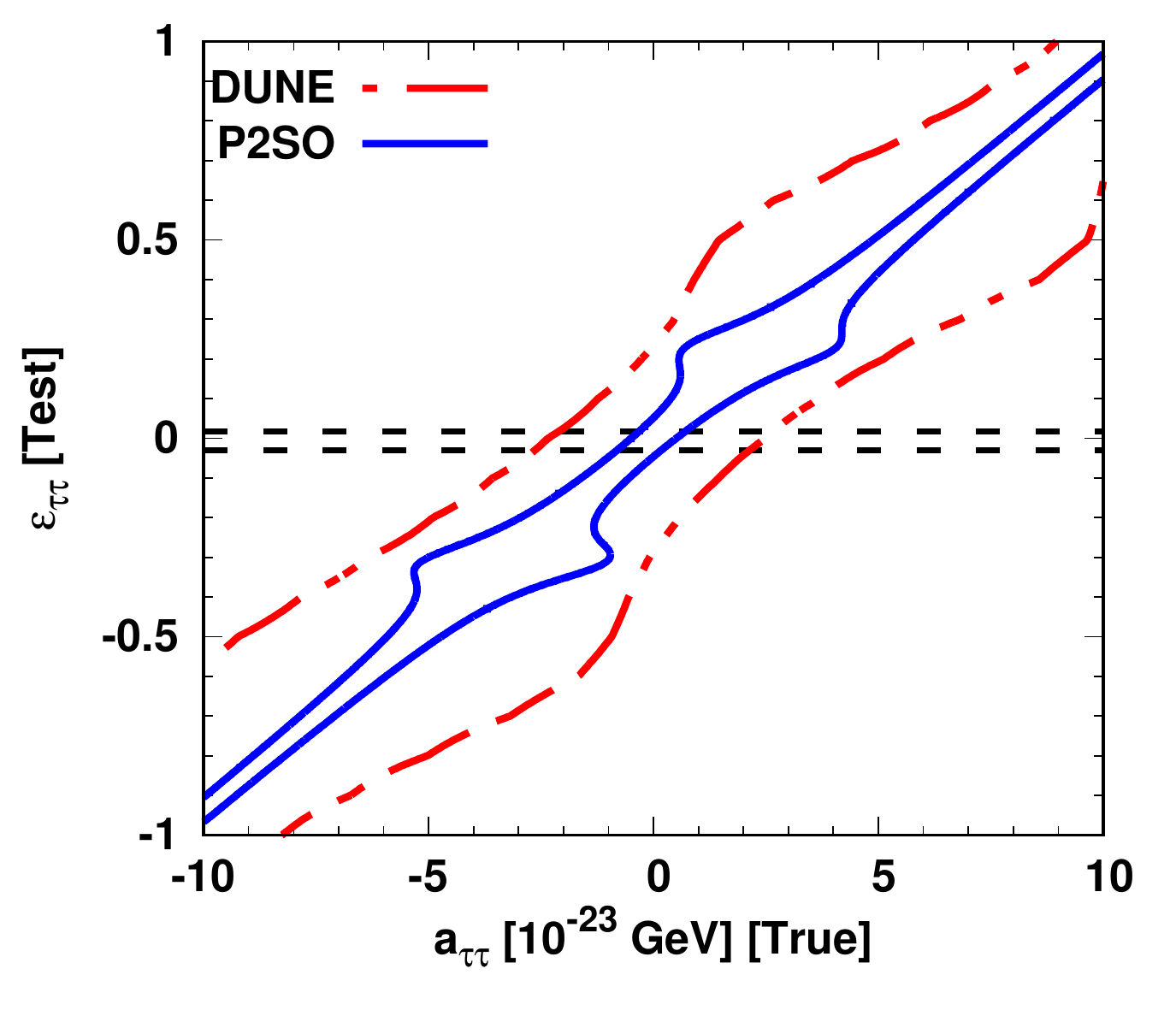}
    \caption{$3 \sigma$ contours in $a_{\alpha \beta}$[True] - $\epsilon_{\alpha \beta}$[Test] plane. Horizontal lines represent the future bounds of corresponding NSI parameters $\epsilon_{\alpha\beta}$.}
    \label{fig:allowed region}
\end{figure*}

To understand the results in a better way, in Fig.~\ref{fig:allowed region}, we examined the same situation in a different way. This figure depicts the contours when we consider LIV in true and test this theory against NSI hypothesis using its future range at $3 \sigma$ C.L. The red and blue contours are for DUNE and P2SO respectively. The horizontal black dashed lines represent the future limits of NSI parameters. The top row is for $e \mu$, $e \tau$ and $e e$ components and the bottom row is for $\mu \mu$, $\mu \tau$ and $\tau \tau$ components. 

The area inside the contours represent the values of the LIV parameters which can't be distinguished from the NSI parameters at $3\sigma$ level. The region outside the contours allow us to discriminate NSI from LIV at the same or higher confidence level. From these panels,  we can also see that in general, the sensitivity is higher for P2SO than DUNE except the $e\mu$ sector. To have a clear understanding of the values of LIV parameters for which we can separate them from NSI, we proceed as follows. If the latter values are constrained by the future experiments, we draw vertical grey lines in the top left panel such that the left vertical line intersects the lower future  bound of NSI and the P2SO curve, while the right vertical line intersects the upper future bound of NSI and the P2SO curve. Therefore, for the values of $a_{e\mu}$ which lie outside the grey vertical lines, it is possible to have $3 \sigma$ separation with $\epsilon_{e\mu}$, considering $\epsilon_{e \mu}$ is constrained by the future experiments. Similar exercise can be done for the other parameters too. We have checked that conclusions obtained in these panels are consistent with Figs. \ref{fig:ee-emu-etau} and \ref{fig:mumu-mutau-tautau} and therefore with Tab.~\ref{res}.

\section{Summary}

In this paper, we made an attempt to discriminate NSI and LIV in the context of the long-baseline experiments DUNE and P2SO. Though the origin of the NSI and LIV theories are very different, they manifest themselves in a similar fashion in the Hamiltonian of the neutrino oscillation. The only difference is that the phenomenon of NSI depends on the earth matter density whereas LIV is independent of the matter density. Therefore, in the fixed baseline experiments where the matter density is constant, it is very difficult to distinguish NSI from LIV, as these effects become almost identical. However, as the present and future bounds of the NSI and LIV parameters are not equivalent, it is possible to separate these two theories with respect to their present/future bounds in the long-baseline experiments. Our choice of the experiments are motivated by the fact that DUNE and P2SO are the two long-baseline experiments which can have most probable sensitivity  to matter effect with highest possible statistics. 

In our work, we demonstrated the possibilities of discriminating LIV from NSI taking the present and future bounds of the NSI parameters. While compiling the bounds on the NSI and LIV parameters, we found that the future bounds of the NSI and LIV parameters are one order of magnitude better than the present available bounds except the parameters $\epsilon_{\mu\mu}$ and $a_{\mu\tau}$.  At the probability level, we have seen that the separation between LIV and NSI is better when one considers the future bounds on the NSI parameters  compared to their present bounds. Considering LIV in the data and NSI in theory, we find that indeed the capability of DUNE and P2SO to distinguish LIV from NSI improves when one considers the future bounds of the NSI parameters  compared to present bounds. For the parameters $ee$ and $\tau\tau$, it is impossible to separate LIV from NSI for the current bounds of NSI parameters. The improvement in the sensitivity due to the future bounds on NSI is higher
in the parameters that appear in the appearance channel, compared for the ones
in the disappearance channel. Between the two experiments, the sensitivity of P2SO is better than DUNE except $e \mu$ sector. Our results show that the best discrimination between LIV and NSI is possible at $3 \sigma$ C.L for the parameter $a_{\mu\mu}$. In this case, the value of the LIV parameter falls inside its future bound, if one considers the value of the NSI parameter to be constrained by the present experiments. The worst sensitivity comes from the sectors $ee$ and $\tau\tau$. For these parameters, one can not separate LIV from NSI, if one considers the NSI parameters to be constrained by the present experiments.

To conclude, we find that the P2SO experiment is best suited to discriminate between the NSI and LIV scenarios and in particular, the new  physics effects associated with the $\mu \mu$ sector. 

\section*{Acknowledgements}

MG would like to thank Leon Halic for useful discussion. DKS acknowledges Prime Minister's Research Fellowship, Govt. of India. RM acknowledges the support from University of Hyderabad through the IoE project grant IoE/RC1/RC1-20-012. We acknowledge the use of CMSD HPC facility of Univ. of Hyderabad to carry out computations in this work. This work has been in part funded by Ministry of Science and Education of Republic of Croatia grant No. KK.01.1.1.01.0001.

\bibliography{NSI-LIV-ref.bib}
\bibliographystyle{JHEP}

\end{document}